\documentclass{article}
\usepackage{dcase2021_techrep,amsmath,graphicx,url,times,booktabs, tabularx}
\usepackage{multirow}
\usepackage{amsmath}
\usepackage{amsfonts}

\title{AUDIO CAPTIONING USING SOUND EVENT DETECTION}

\twoauthors
{Ayşegül Özkaya Eren}
{Başkent University\\
	Department of Computer Engineering\\
	Ankara, Turkey \\
	21610279@mail.baskent.edu.tr}
{Mustafa Sert}
{ Başkent University \\
	Department of Computer Engineering\\
	Ankara, Turkey \\
	msert@baskent.edu.tr}

\begin{document}
	
	\ninept
	\maketitle
	
\begin{sloppy}
\begin{abstract}
This technical report proposes an audio captioning system for DCASE 2021 Task 6 audio captioning challenge. Our proposed model is based on an encoder-decoder architecture with bi-directional Gated Recurrent Units (BiGRU) using pretrained audio features and sound event detection. A pretrained neural network (PANN) is used to extract audio features and Word2Vec is selected with the aim of extracting word embeddings from the audio captions. To create semantically meaningful captions, we extract sound events from the audio clips and feed the encoder-decoder architecture with sound events in addition to PANNs features. Our experiments on the Clotho dataset show that our proposed method significantly achieves better results than the challenge baseline model across all evaluation metrics.
		\end{abstract}
		
		
		\section{Introduction}
		\label{sec:intro}
		
The automated audio captioning task is a combination of audio and natural language processing to create meaningful natural language sentences \cite{DBLP:journals/corr/DrossosAV17}. The purpose of audio captioning is different than the previous audio processing tasks such as audio event/scene detection and audio tagging. The previous tasks do not aim to create descriptive natural language sentences whereas audio captioning aims to capture relations between events, scenes, and objects to create meaningful sentences. This technical report presents the details of our submission for DCASE 2021 Task 6 automated audio captioning. We propose an encoder-decoder model using sound event detection and PANNs (Pretrained Audio Neural Networks).

Since sound events in audio clips are informative to capture the main context of an audio clip, we propose a new model using sound event detection to obtain more semantic information.
		
This paper is organized as follows: Section 2 describes the details of system architecture. The experimental setup and dataset information are presented in Section 3. Section 4 shows the results and Section 5 gives the conclusion.

\section{SYSTEM ARCHITECTURE}
\label{sec:system} 
The overall system architecture is shown in Figure 1. The details of the proposed model are presented in this section.

\begin{figure*}
	\centering	
	\includegraphics[scale=0.5]{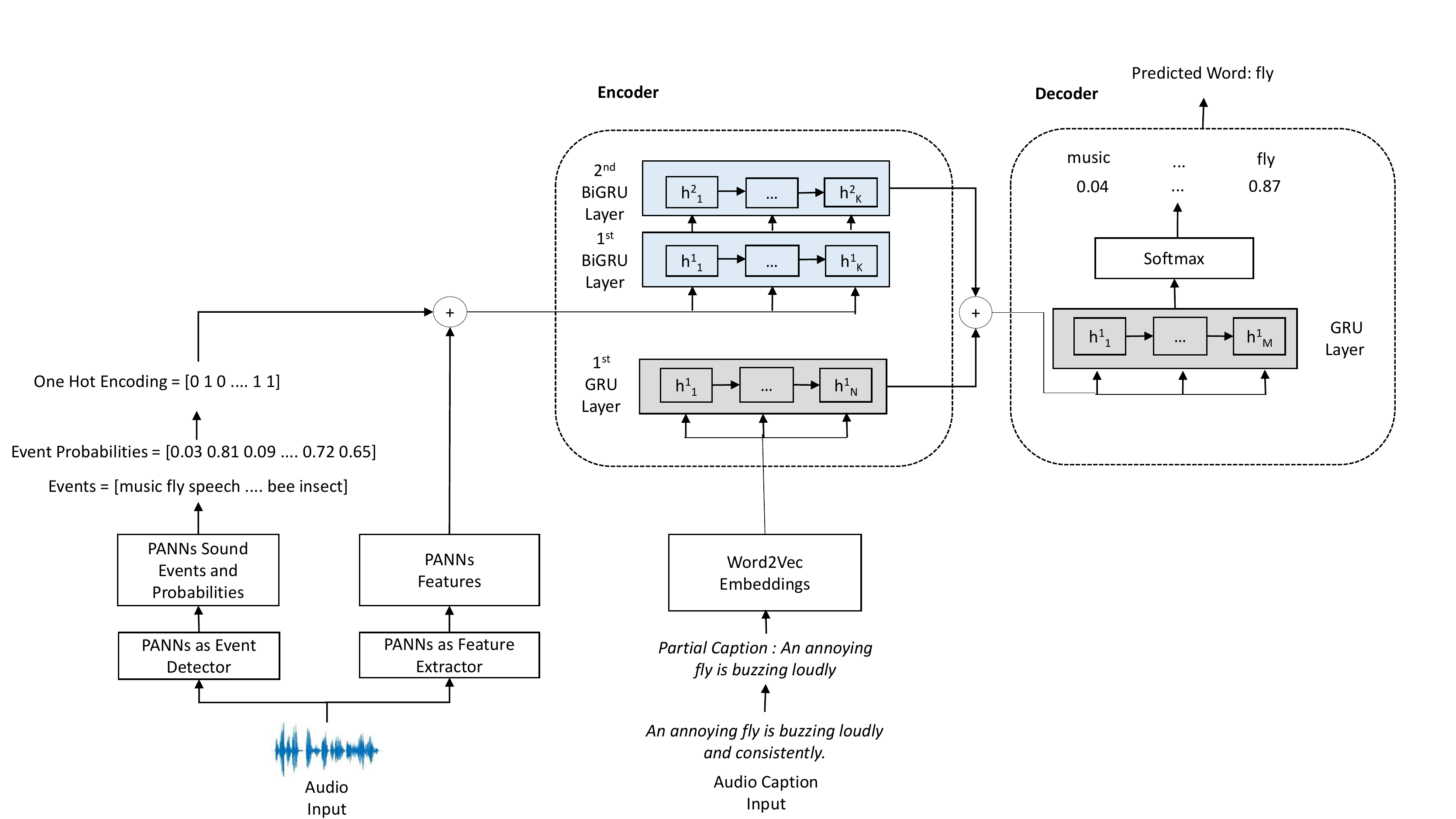}
	\caption{The illustration of the proposed audio captioning model. The PANNs used to extract both audio features and sound events. (+) is used for concatenation method.}
	\label{figModelPng}
\end{figure*}

\subsection{PANNs Feature Extraction}
\label{ssec:panns}
The PANNs are pretrained features on the AudioSet dataset \cite{7952261}. Wavegram-Logmel-CNN14 model is used to extract the PANNs features. 96 ms Hamming window and 50\% overlap are applied for windowing and overlapping methods similar to \cite{Drossos_2020}. We present $\textbf{x}=[x_1,...,x_T], T \in \mathbb{R}^{2048}$ as PANNs features.

\subsection{Sound Event Extraction}
\label{ssec:sound}

The PANNs are used to extract sound events. The last layer of the PANNs gives the probabilities of each sound event on the AudioSet dataset. The dataset contains 527 sound classes. We obtain $\textbf{e}=[e_1,...,e_M], e_m \in \mathbb{R}^{527}$, where $e_m$ is the probability of each sound classes on the AudioSet dataset. 

We apply a 0.1 threshold value to the sound event probabilities and the events greater than 0.1 probability are selected for each audio clip. Therefore the most probable events are obtained for a given audio clip. After that, an event tokenizer is generated from the AudioSet events to divide some events that have more than one word. The purpose of tokenization is to obtain the similarity of words in different sound events. For instance, the AudioSet contains different classes such as ``{\textit{Funny Music}}'', ``{\textit{Sad Music}}'', ``{\textit{Scary Music}}'', ``{\textit{Middle Eastern Music}}'' etc. The tokenization method can capture the similarities between four audio clips that contain different music events. After the tokenization, we obtain an event corpus represented as $\textbf{c}=[c_1,...,c_K], c_K \in \mathbb{R}^{600}$. Then, the sound event vector of each audio clip is obtained by using one-hot-encoding.  For the $j^{th}$ audio clip,  if the audio clip contains $c^{th}$ event, $c_{ik}$=1 otherwise $c_{ik}$=0.

\subsection{Encoder-Decoder}
\label{ssec:encoder}
We use the same encoder-decoder model in \cite{9327916}. In the proposed encoder-decoder architecture, there are two BiGRU layers for encoding PANNs features. The first BiGRU layer contains 32 cells and the second BiGRU layer contains 64 cells. There is one GRU layer that contains 128 cells for encoding partial captions. The number of the cells is selected empirically. After obtaining a sound event vector for each audio file, PANNs features and sound event vectors are concatenated as an input to the encoder. 

The captions in the dataset are preprocessed before feeding the model. All words are converted to lowercase and all punctuations are removed. \texttt{$<$sos$>$} and \texttt{$<$eos$>$} are added to the beginning and end of the captions. Previous studies show that the inclusion of Word2Vec \cite{DBLP:journals/corr/MikolovSCCD13} improves the performance of the audio captioning system \cite{eren2020audio}, therefore Word2Vec is used for representing the captions in the training dataset. Each unique word in the dataset is represented by $v=[v_1,...,v_i]$, where $v_i\in \mathbb{R}^{256}$ and 256 is the dimension for word embeddings. Encoded partial captions are concatenated to the encoded PANNs features and sound event vectors to feed the decoder part. 

The decoder contains one GRU layer that contains 128 cells. The Softmax function is used after the fully connected layer. The decoder predicts probabilities of the unique words in the dataset and selects the most probable word as the predicted word. After finding the \texttt{$<$eos$>$} token, the whole sentence is created by the predicted words.

\section{EXPERIMENTS}
\label{sec:experiments}
This section describes the details of the dataset and implementation details.

\subsection{Dataset}
\label{ssec:dataset}
We use Clotho \cite{Drossos_2020} audio captioning dataset for our experiments. The challenge presents a new version of the Clotho dataset that is called Clotho V2. Clotho V2 contains 3840 audio clips in the development split. There is a validation split that has 1046  audio files in the Clotho V2. There are 1045 audio files in the evaluation split and 1043 audio files in the test split. All of the splits have five captions for each audio clip. For our experiments, we use five times each audio file with their corresponding captions similarly to \cite{Drossos_2020}.

\subsection{Experimental Setup}
\label{ssec:setup}
Our system is implemented using Keras framework \cite{keras} and our experiments are run on a computer with GPU GTX1660Ti, Linux Ubuntu 18.04 system. Python 3.6 is used for implementation. We run all experiments for 100 epochs and we choose the model with the minimum validation error. Different batch sizes are applied to our proposed model with the best results. 128 is chosen as batch-size. Adam optimizer, LeakyReLU activation function, and cross-entropy loss are used as hyperparameters. Batch normalization \cite{DBLP:journals/corr/IoffeS15} and a dropout rate of 0.5 are also used. The number of parameters on our prosed model is nearly 2,500,000.

\section{RESULTS}
\label{sec:results}

For evaluation, BLUE-n \cite{Papineni2002}, METEOR \cite {Banerjee2005}, ROUGE$_L$ \cite{Lin2004}, CIDEr \cite{Vedantam2015}, SPICE \cite{10.1007/978-3-319-46454-1_24}, and SPIDEr \cite {Liu_2017} metrics are used. The matching words in the actual and predicted captions are calculated for BLEU-n. It calculates the precision for n-grams. Recall and precision are calculated for METEOR. ROUGE$_L$ calculates Longest Common Subsequence. CIDEr presents more semantic results by calculating cosine-similarity between actual and predicted captions. SPICE parses the actual and predicted captions and creates scene graph representation. SPIDEr is a linear combination of CIDEr and SPICE.

The comparison of our proposed method and Clotho V2 is shown in Table 1. The results show that our model significantly outperforms the challenge baseline across all evaluation metrics. 

 \begin{table*}	[t]
	\caption{The comparison of our proposed method and Clotho V2 baseline results} 
\begin{center}
	\resizebox{\textwidth}{!}{
		\begin{tabular}{ l|l|c|c|c|c|c|c|c|c}
			\hline
			\multirow{2}*{\bfseries Method} & \multicolumn{9} {c}{\bfseries Metric} \\
			\cline{2-10}
			& \textbf{BLEU-1}& \textbf{BLEU-2}& \textbf{BLEU-3} & \textbf{BLEU-4} &  \textbf{METEOR} & \textbf{ROUGE$_L$} & \textbf{CIDEr} & \textbf{SPICE} & \textbf{SPIDEr}\\
			
			\hline
			\textbf{Clotho V2 baseline}    &       0.378     &   0.119 &  0.050      &    0.017   &    0.078 &    0.263 &    0.075 & 0.028 & 0.051\\
			
			\textbf{Proposed Method}    &      \textbf{0.586}    &  \textbf{0.356} &   \textbf{0.268}      &     \textbf{0.150}  &    \textbf{0.214} &    \textbf{0.444} &    \textbf{0.328} & \textbf{0.155} & \textbf{0.242} \\
			
			\hline
		\end{tabular}
	}
\end{center}
\label{table1}
\end{table*}

The predicted captions on the evaluation dataset show that our proposed model can predict meaningful captions. 
\section{CONCLUSION}
\label{sec:conclusion}
This technical report presents our system details for participating DCASE 2021 Task 6. An encoder-decoder model with sound event detection and pretrained audio features is proposed for the challenge. The results show that the inclusion of sound event detection improves the audio captioning performance. In future work, different fusion and extraction methods will be applied to the audio features and sound events.

\bibliographystyle{IEEEtran}
\bibliography{refs}

\begin{thebibliography}{10}
\providecommand{\url}[1]{#1}
\def\UrlFont{\rmfamily}
\providecommand{\newblock}{\relax}
\providecommand{\bibinfo}[2]{#2}
\providecommand\BIBentrySTDinterwordspacing{\spaceskip=0pt\relax}
\providecommand\BIBentryALTinterwordstretchfactor{4}
\providecommand\BIBentryALTinterwordspacing{\spaceskip=\fontdimen2\font plus
\BIBentryALTinterwordstretchfactor\fontdimen3\font minus
  \fontdimen4\font\relax}
\providecommand\BIBforeignlanguage[2]{{%
\expandafter\ifx\csname l@#1\endcsname\relax
\typeout{** WARNING: IEEEtran.bst: No hyphenation pattern has been}%
\typeout{** loaded for the language `#1'. Using the pattern for}%
\typeout{** the default language instead.}%
\else
\language=\csname l@#1\endcsname
\fi
#2}}

\bibitem{DBLP:journals/corr/DrossosAV17}
K.~{Drossos}, S.~{Adavanne}, and T.~{Virtanen}, ``Automated audio captioning
  with recurrent neural networks,'' in \emph{2017 IEEE Workshop on Applications
  of Signal Processing to Audio and Acoustics (WASPAA)}, 2017, pp. 374--378.

\bibitem{7952261}
J.~F. {Gemmeke}, D.~P.~W. {Ellis}, D.~{Freedman}, A.~{Jansen}, W.~{Lawrence},
  R.~C. {Moore}, M.~{Plakal}, and M.~{Ritter}, ``Audio set: An ontology and
  human-labeled dataset for audio events,'' in \emph{2017 IEEE International
  Conference on Acoustics, Speech and Signal Processing (ICASSP)}, 2017, pp.
  776--780.

\bibitem{Drossos_2020}
K.~Drossos, S.~Lipping, and T.~Virtanen, ``Clotho: An audio captioning
  dataset,'' in \emph{ICASSP 2020-2020 IEEE International Conference on
  Acoustics, Speech and Signal Processing (ICASSP)}.\hskip 1em plus 0.5em minus
  0.4em\relax IEEE, 2020, pp. 736--740.

\bibitem{9327916}
A.~{Özkaya Eren} and M.~{Sert}, ``Audio captioning based on combined audio and
  semantic embeddings,'' in \emph{2020 IEEE International Symposium on
  Multimedia (ISM)}, 2020, pp. 41--48.

\bibitem{DBLP:journals/corr/MikolovSCCD13}
\BIBentryALTinterwordspacing
T.~Mikolov, I.~Sutskever, K.~Chen, G.~Corrado, and J.~Dean, ``Distributed
  representations of words and phrases and their compositionality,''
  \emph{CoRR}, vol. abs/1310.4546, 2013. [Online]. Available:
  \url{http://arxiv.org/abs/1310.4546}
\BIBentrySTDinterwordspacing

\bibitem{eren2020audio}
A.~{\"O}. Eren and M.~Sert, ``Audio captioning using gated recurrent units,''
  \emph{arXiv preprint arXiv:2006.03391}, 2020.

\bibitem{keras}
F.~Chollet \emph{et~al.}, ``Keras,'' \url{https://github.com/fchollet/keras},
  2015.

\bibitem{DBLP:journals/corr/IoffeS15}
\BIBentryALTinterwordspacing
S.~Ioffe and C.~Szegedy, ``Batch normalization: Accelerating deep network
  training by reducing internal covariate shift,'' \emph{CoRR}, vol.
  abs/1502.03167, 2015. [Online]. Available:
  \url{http://arxiv.org/abs/1502.03167}
\BIBentrySTDinterwordspacing

\bibitem{Papineni2002}
K.~Papineni, S.~Roukos, T.~Ward, and W.-j. Zhu, ``{BLEU : a Method for
  Automatic Evaluation of Machine Translation},'' \emph{Computational
  Linguistics}, no. July, pp. 311--318, 2002.

\bibitem{Banerjee2005}
S.~Banerjee and A.~Lavie, ``{METEOR: An automatic metric for MT evaluation with
  improved correlation with human judgments},'' \emph{Proceedings of the acl
  workshop on intrinsic and extrinsic evaluation measures for machine
  translation and/or summarization}, vol.~29, pp. 65--72, 2005.

\bibitem{Lin2004}
C.~Y. Lin, ``{Rouge: A package for automatic evaluation of summaries},''
  \emph{Proceedings of the workshop on text summarization branches out (WAS
  2004)}, no.~1, pp. 25--26, 2004.

\bibitem{Vedantam2015}
R.~Vedantam, C.~L. Zitnick, and D.~Parikh, ``{CIDEr: Consensus-based image
  description evaluation},'' \emph{Proceedings of the IEEE Computer Society
  Conference on Computer Vision and Pattern Recognition}, vol. 07-12-June-2015,
  pp. 4566--4575, 2015.

\bibitem{10.1007/978-3-319-46454-1_24}
P.~Anderson, B.~Fernando, M.~Johnson, and S.~Gould, ``Spice: Semantic
  propositional image caption evaluation,'' in \emph{Computer Vision -- ECCV
  2016}, B.~Leibe, J.~Matas, N.~Sebe, and M.~Welling, Eds.\hskip 1em plus 0.5em
  minus 0.4em\relax Cham: Springer International Publishing, 2016, pp.
  382--398.

\bibitem{Liu_2017}
\BIBentryALTinterwordspacing
S.~Liu, Z.~Zhu, N.~Ye, S.~Guadarrama, and K.~Murphy, ``Improved image
  captioning via policy gradient optimization of spider,'' \emph{2017 IEEE
  International Conference on Computer Vision (ICCV)}, Oct 2017. [Online].
  Available: \url{http://dx.doi.org/10.1109/ICCV.2017.100}
\BIBentrySTDinterwordspacing

\end{thebibliography}
%
%
%
%
%
%
%
%
%

\end{sloppy}
\end{document}